# Geology and Photometric Variation of Solar System Bodies with Minor Atmospheres: Implications for Solid Exoplanets

Yuka Fujii[*1], Jun Kimura[1], James Dohm[2], and Makiko Ohtake[3]

[1]Earth-Life Science Institute, Tokyo Institute of Technology, Tokyo, Japan

[2]The University Museum, The University of Tokyo, Tokyo, Japan

[3]Institute of Space and Astronautical Science, Japan Aerospace Exploration Agency, Kanagawa, Japan

*yuka.fujii@elsi.jp

TIME STAMP:

**ABSTRACT**

A reasonable basis for future astronomical investigations of exoplanets lies in our best knowledge of the planets and satellites in the Solar System. Solar System bodies exhibit a wide variety of surface environments, even including potential habitable conditions beyond Earth, and it is essential to know how they can be characterized from outside the Solar System. In this study, we provide an overview of geological features of major Solar System solid bodies with minor atmospheres (i.e., the Terrestrial Moon, Mercury, the Galilean moons, and Mars) that affect surface albedo at local to global scale, and we survey how they influence point-source photometry in UV/visible/near IR (i.e., the reflection dominant range). We simulate them based on recent mapping products and also compile observed light curves where available. We show a 5–50% peak-to-trough variation amplitude in one spin rotation associated with various geological processes including heterogeneous surface compositions due to igneous activities, interaction with surrounding energetic particles, and distribution of grained materials. Some indications of these processes are provided by the amplitude and wavelength dependence of variation in combinations of the time-averaged spectra. We also estimate the photometric precision needed to detect their spin rotation rates through periodogram analysis. Our survey illustrates realistic possibilities for inferring the detailed properties of solid exoplanets with future direct imaging observations.

Keywords:   Planetary environments—Planetary geology—Solar System—Extrasolar terrestrial planets.

## 1. INTRODUCTION

Direct imaging of rocky planets is expected to play an essential role in looking for signatures of life and investigating the surface environments of exoplanets. To develop our ability to decipher the light from unknown exoplanets, one possible starting point is to understand the photometric and spectroscopic features of well-known Solar System planets/satellites as if they were extrasolar bodies. For this purpose, Earth, among others, has been of significant interest in terms of habitability, and its disk-integrated spectrum has been studied both observationally (e.g., Arnold et al. 2002; Woolf et al. 2002; Hamdani et al. 2006; Turnbull et al. 2006; Livengood et al. 2011) and theoretically (e.g., Ford et al. 2001; Des Marais et al. 2002; Tinetti et al. 2006a,b; Robinson et al. 2011; Robinson 2011; Kaltenegger et al. 2010). Spectra of the Earth at different evolutionary stages have been also modeled by changing the bioactivity (Kaltenegger et al. 2007; Sanromá et al. 2013) and continental distribution (Sanromá & Pallé 2012).

The appearance of other Solar System bodies also needs to be considered[1] as more general samples of exoplanets beyond Earth, because they are the only samples of planets whose real nature can be known in detail. In reality, various geological processes alter the surface reflectance of Solar System solid bodies at local to global scale, and they provide clues to the current conditions and evolution histories on which planetary habitability relies. Therefore, it is important to examine the detectability of their surface signatures from a distance. In the context of future applications to exoplanets, spectra of Solar System bodies have been compiled and characterization based on color-color plots has been discussed (Lundock et al. 2009; Crow et al. 2011; Traub 2003). Mallama (2009) proposed the characterization of terrestrial exoplanets through orbital phase curves based on the empirical phase curves of Solar System planets. Hu et al. (2012) studied spectroscopic signatures of theoretical planets with various surface compositions inspired by Solar System bodies and put forward several observable features including Si-O features in the mid-infrared, an Fe band around 1 μm, and O-H bands in the near IR (NIR).

Along with these properties, time variability due to spin rotation could provide complementary information about surface geology. In the case of Earth, disk-averaged scattered light exhibits diurnal variation coherent with the change in the composition in the illuminated and visible region. Because the continental distribution is highly inhomogeneous, and the global cloud pattern does not change significantly in a day, the spin rotation period may be successfully identified through periodogram analysis (Pallé et al. 2008), which then may allow for recovery of surface inhomogeneity along the equator (Cowan & Agol 2008; Cowan et al. 2009, 2011; Oakley & Cash 2009; Fujii et al. 2010, 2011, 2013) as well as estimate the surface composition (Cowan & Strait 2013).

In this study, we surveyed photometric variations of Solar System bodies with minor atmosphere and their relationship to surface geology by performing simulations based on the latest mapping products and by adopting the observed data of multi-band photometry. In particular, the Terrestrial Moon, Mercury, the Galilean moons, and Mars are considered[2]. Our intent was to study the general causes of photometric variation, estimate variation amplitude, and explore the possibility of identifying geological features as well as the spin rotation periods. After reviewing our current understandings of the surface colors and its relation to geological characteristics of Solar System bodies (Section 2), we describe our methods and datasets used to obtain multi-band light curves for each body (Section 3). In Section 4, we present our main results, including simulated light curves, average spectra, and wavelength-dependence of variation, and compare them with their characteristic geologies. The detectability of the spin rotation period is also examined in Section 4. In Section 5, we discuss the effects of atmospheres on photometric characterization; present another noteworthy example from the minor bodies in the Solar System, Iapetus; and offer a general caution on the interpretations of the photometric properties. Our conclusions are summarized in Section 6.

---
[1] We use the word "body" to refer to a planet or a moon without distinguishing them.
[2] We include Mars, although containing a minor atmosphere of ~7.5 mbar, which influences its appearance; the surface of Mars is directly observable except during dust storms compared to the cloud(haze)-shrouded Venus and Titan with surface colors completely obscured.

## 2. BRIEF REVIEW OF THE SURFACE GEOLOGICAL FEATURES OF SOLAR SYSTEM BODIES WITH MINOR ATMOSPHERES

In this section, we provide an overview of the current understandings about the surface geologies of major Solar System solid bodies with minor atmospheres. We focus on features that influence planetary albedos in UV/visible/NIR because they are parameters that are primarily accessible by future direct imaging observations of exoplanets.

In short, several major processes are identified as factors that vary planetary albedo regionally. These include compositional diversity due to igneous activities (e.g., on Io and the Terrestrial Moon), interactions between surface ice and surrounding plasma particles (on Europa, Ganymede, and Callisto), global tectonic deformation (e.g., on Ganymede), and overlay of fine-grained materials (on Mars).

Below, we discuss the albedo characteristics of each body and their geological origins in detail.

### 2.1. Terrestrial Moon

The Terrestrial Moon is well known for the prominent dichotomy in its surface albedo. Bright areas (highlands) are predominantly made of anorthosite, which were deposited from a magma ocean in the very early stage of Lunar history (Taylor 1982; Warren 1990). On the other hand, the dark areas (maria) are large impact basins that were filled with mafic basaltic lavas through volcanic activity in the later stage (the estimated age of mare basalt ranges from ~4 Ga to 1.5 Ga; Morota et al. 2011). The greatest extent of maria is on the near side (Earth-facing hemisphere) (Wilhelms et al. 1987), leaving the far side brighter. This hemispheric asymmetry is likely related to the difference in crustal thickness between the near and far sides; an equipotential surface is closer to the surface of the near side, and thus, magmas that originate at the same level can reach the surface more easily than on the far side (e.g., Hiesinger & Head 2006). Although the origin of the dichotomy in crustal thickness is still controversial, one convincing hypothesis is that it is due to massive impact events (Cadogan 1981; Whitaker 1981; Nakamura et al. 2012). Typical examples of the reflectance spectra of highland and mare are shown in the left panel of Fig. 1. They exhibit a remarkable contrast in amplitude.

**2.2. Mercury**

Mercury exhibits relatively homogeneous gray color and is known to be covered with basaltic materials with common sulfur- and magnesium-rich characteristics (Nittler et al., 2011, Rhodes et al., 2011). At a closer look, there are both smooth plains and heavily cratered regions. Smooth plains, comprising 40% of the surface, are thought to be volcanic plains erupted through fractured crust contemporaneous with the period of Late Heavy Bombardment[3] (LHB; 4.1 to 3.8 Ga), while the heavily cratered regions indicate ancient crust. The color contrast among different geological units are much less obvious than for the Lunar surface partly because the colors of the volcanic deposits are brighter than their Lunar counterpart (i.e., mare basalt) due to the lower amount of iron (Nittler et al. 2011; Rhodes et al. 2011; Lucey et al. 2006). In addition, geological activity on Mercury has been apparently dormant since the LHB, leaving a large percentage of the surface ancient (>3.8 Ga) (Strom 1979) and saturated with impact craters; without geological activity, the long-lasting impact cratering process can deform old terrains and mix the volcanic materials (e.g. Arvidson et al. 1975, Schultz 1989), which may have resulted in the relatively homogeneous color of Mercury's surface.

**2.3. Galilean Moons**

The Galilean moons (Io, Europa, Ganymede, and Callisto) are the four largest moons that orbit Jupiter in synchronous rotation states (i.e., for each body, the spin rotation period and orbital period coincide), implying one side permanently faces the planet, and a permanent "leading" and "trailing" hemisphere with respect to the orbital plane. The surface geologies of the Galilean moons are strongly affected by tidal forces due to the gravitational interaction with Jupiter, as well as by the interaction with surrounding plasmas that are produced through the acceleration of Io-derived particles by the fast-rotating Jovian magnetosphere.

All the moons orbiting Jupiter, except Io, have icy surfaces. While surfaces covered with new, pure, and fine $H_2O$ particles are typically bright and the albedo can reach close to 1, the actual geometric albedos of Europa, Ganymede, and Callisto are ~0.7 (Moore et al. 2004), ~0.4 (Morrison & Morrison 1977), and ~0.2 (Moore et al. 2004), respectively. Surface darkening likely originates from the following two processes. First, the longer the icy surface is exposed to space, the more it is contaminated by meteoritic materials; hydrated non-ice materials are generally dark and exhibit weak ultraviolet $Fe^{3+}$ absorption and a positive visible-to-NIR slope, similar to some carbonaceous chondrites (space weathering) (Clark 1980). Second, the annealing of ice results in larger grains and, hence, a darker surface (Johnson 1997). In turn, brighter icy surfaces are generally younger, indicating the presence of dynamical activities such as plumes (e.g., Porco et al. 2006) and/or cryovolcanism (e.g., Pappalardo et al. 1998). Because these activities can be maintained by internal heat sources and the consequent mobility of the interior, the brighter surfaces may suggest the possibility of subsurface liquid layers, which could provide a potential habitat for extraterrestrial life.

A common albedo feature among Europa, Ganymede, and Callisto is the contrast between the leading and trailing hemispheres (e.g., Stebbins & Jacobsen 1928; Harris 1961; Johnson 1971; Blanco & Catalano 1974; Morrison et al. 1974; Millis & Thompson 1975; Nash & Johnson 1979). This pattern has been discussed in relation to the fact that the energetic plasmas trapped in the Jovian magnetosphere preferentially hit the trailing hemispheres, although the details of the interactions between the surfaces and energetic particles remain uncertain. The degree of the hemispheric albedo dichotomy depends on the effects of excavation of pristine ice by meteoroid impacts (more influential on leading hemispheres) and radiolytic darkening processes (more influential on trailing hemispheres). Similar processes would work on synchronously rotating body surrounded by plasmas or other materials.

---

[3]The Late Heavy Bombardment, also referred to as the Lunar cataclysm, is a putative event thought to have occurred approximately 4.1 to 3.8 Ga during which a large number of impact craters may have formed on the Moon and on other terrestrial planets.

### 2.3.1. Io

The surface of Io, the innermost of the four Galilean moons, shows diverse visible colors, which are identified as various volcanic-related sulfur-containing materials, such as yellow (sulfur), brown (radiolytically decomposed sulfur chains, $S_x$), gray ($SO_2$ frost), and black (silicate pyroclastics) (Carlson et al. 2007 and references therein). Indeed, Io has the most intense volcanic activity in the Solar System, and more than 150 active volcanoes have been confirmed and distributed across the globe. These activities are sustained by the strong tidal interaction with Jupiter through deformation and dissipated heat.

### 2.3.2. Europa

The surface of Europa is mostly covered with bright ice, as a consequence of the active resurfacing process maintained by internal heat sources; the age of Europa's surface is estimated at ~20-180 Ma (Zahnle et al. 2003). However, there is a broad, brownish area in the trailing hemisphere. This is likely due to hydrated sulfuric acid, which can be naturally produced from the radiolysis of Io-derived sulfurous material and surface $H_2O$ (Carlson et al. 2002).

In addition, Europa's surface shows global cracked and disrupted terrains with dark colors. The NIR mapping spectrometer onboard the Galileo spacecraft has identified some hydrated salt minerals (e.g., magnesium, sodium sulfate, and/or carbonate) that are strongly associated with geological processes (McCord et al. 1998). While most hydrated salts are colorless, radiolytic products of sulfate and possibly sulfide compounds—in particular polymeric sulfur (e.g., $S_4$ and $S_8$)—serve as coloring agents.

### 2.3.3. Ganymede

The surface geology of Ganymede is classified into two units. Forty percent of the surface is old and dark (~4.0 Gyr; Zahnle et al. 2003), while the remaining 60% is relatively young and brighter (~2.0 Gyr; Zahnle et al. 2003). The younger and brighter regions contain evidence of global tectonic deformation such as extensive arrays of grooves and ridges. Currently, it is believed that the lithosphere is extensionally strained because of global expansion (Showman et al. 1997) possibly caused by past internal heating events. In addition, Ganymede has polar caps (greater than ~40 deg) that are likely composed of water frost and are noticeably brighter than the equatorial regions (Khurana et al. 2007).

### 2.3.4. Callisto

The surface of Callisto is globally dark, probably due to long exposure to space without resurfacing. The age of Callisto's surface is estimated to be ancient (~4.0 Gyr; Schenk et al. 2004) with no evidence of tectonic activity, probably because the tidal force exerted on Callisto in its distant orbit from Jupiter is not enough to produce sufficient heat. Callisto also shows albedo contrast between leading and trailing hemispheres, but it depends on the geometric configuration between the Sun, Callisto, and the observer. Except for opposition phase, that is, at the configuration where the direction of the Sun and the observer coincide, the trailing hemisphere is brighter in visible light (e.g., Moore et al. 2004), unlike Europa and Ganymede. Near the opposition phase, the leading hemisphere is brightened up ("opposition surge") more prominently than the trailing hemisphere resulting in even reversed contrast, possibly because of the rougher and less-compacted leading hemisphere due to the enhanced micrometeoritic erosion (e.g., Buratti 1995).

### 2.4. Mars

The martian surface exhibits a global red color due to iron oxides ("rust"), the origin of which is still controversial. Increasing evidence points to aqueous activities in the past (e.g., Head et al. 1999, McEwen et al. 2007, Squyres et al. 2012), which would have contributed to the formation of rust; rust forms when oxygen comes in long-term contact with iron.

In addition, the martian surface shows significant global variation in albedo. Although the origins of the albedo features are not fully understood, bright regions (typically with albedos larger than 0.2) are largely attributed to fine-grained materials (< 10 μm; e.g., Ruff and Christensen 2002) (see the bottom panel of Fig.1), which are mostly created by wind through mechanical weathering; in general, smaller particle size results in higher albedo (e.g., Adams and Filice 1967). Other contributing factors to the albedo signature likely include variations in rock composition affected by hydrological/geological activities through time (e.g., Tharsis-driven transient flooding, ocean formation, and volcanism; Baker et al., 1991, 2007; Dohm et al., 2007, 2009). The albedo distribution pattern has not changed significantly over the past few decades, according to a comparison of the observations by Viking and the Mars Global Surveyor (MGS).

Another notable feature of the martian surface is the hemispheric dichotomy in surface elevation --- the cratered highlands of the southern hemisphere and the lowlands of the northern hemisphere --- the origin of which is still under debate (for details, see Dohm et al. 2013 and references therein). In addition, the Tharsis volcanic bulge that dominates the western hemisphere (centered near the equator and ~265 longitude) is also distinct (Carr and Head, 2010). These factors may also have affected the distribution of grained materials, while there does not appear to be an one-to-one relationship due to the complex interactions of geologic, hydrologic, periglacial, glacial, gravity-driven, and eolian processes through time.

## 3. METHOD AND DATASETS TO COMPILE ROTATIONAL COLOR VARIATION

Motivated by the diversity in the planetary surfaces reviewed in Section 2, we studied the photometric properties of the Solar System bodies as if they were extrasolar objects (i.e., as point sources), with a focus on the time variation according to spin rotation. We did so by running simulations of the disk-integrated light curves based on the pixel-level scattering properties of each body and by collecting the observed disk-integrated photometry of each body. We detail our methods below.

### 3. 1. Simulations

We performed simulations of light curves by the following procedure (see also e.g., Ford et al. 2001, Fujii et al. 2010). First, the relative configuration of the host star (i.e., the Sun), body, and observer is fixed. We assume that the star and observer are in the equatorial plane of the body and that both the star and observer are distant (i.e., the incident light and the scattered light is treated as parallel to the star-body line and body-observer line, respectively). Then, the configuration is specified by the star-body-observer angle ("phase angle") $\alpha$ (left panel of Fig. 2).

Next, the surface of the body is pixelated into 2 deg × 2 deg pixels. The incident zenith angle, $\theta_0$, the zenith angle of observation, $\theta_1$, and the relative azimuthal angle, $\phi$, are computed for each pixel based on the relative configuration of the star and observer as well as the location of the pixel (right panel of Fig. 2). The scattering property of each surface pixel is represented by the bidirectional reflectance distribution function (BRDF), the ratio of the scattered radiance, $I(\theta_0, \theta_1, \phi)$, to the irradiance, $F_0$, $f(\theta_0, \theta_1, \phi; \mathbf{p}_{\vec{r}}) \equiv \frac{I(\theta_0, \theta_1, \phi)}{F_0}$. Here, $\vec{r}$ denotes the location of the pixel, and $\mathbf{p}_{\vec{r}}$ is the parameter set specifying the function whose values vary pixel to pixel.

Finally, the contribution from each surface pixel is calculated by using BRDF with a computed geometric parameter, $\theta_0$, $\theta_1$, and $\phi$, and then the contributions are summed up to obtain the total scattered intensity from the body, as seen as a point source. We neglect the effect of atmospheres.

The necessary information to perform the integral is the BRDF at each pixel, $f(\theta_0, \theta_1, \phi; \mathbf{p}_{\vec{r}})$. We approximately rewrite the BRDF as a product of the reflectance, $a_{\vec{r}}$, and the function responsible for the anisotropy $\tilde{f}(\theta_0, \theta_1, \phi; \mathbf{p})$, that is, $f(\theta_0, \theta_1, \phi; \mathbf{p}_{\vec{r}}) = a_{\vec{r}} \tilde{f}(\theta_0, \theta_1, \phi; \mathbf{p}_{\vec{r}})$. Then, we assign values for $a_{\vec{r}}$ and $\mathbf{p}_{\vec{r}}$ based on the available mapping datasets. The data used in this paper are described below and summarized in Table 1. The examples of the reflectance maps are displayed in Fig. 3.

For the Terrestrial Moon, we obtain reflectance maps at the 414, 749, 901, 950, and 1000 nm bands derived with the Multi-band Imager (MI) onboard the lunar orbiting spacecraft SELENE (Kato et al. 2008) and use them to find $a_{\vec{r}}$ (Fig. 2). We adopt the same anisotropic function, $\tilde{f}(\theta_0, \theta_1, \phi; \mathbf{p})$, as the one used in the derivation of $a_{\vec{r}}$, which is a combination of the phase function and the lunar Lambert function (Eq.[11] of Yokota et al. 2011). Since the function includes four parameters, $\mathbf{p}_{\vec{r}}$, which depend on the reflectance $a_{\vec{r}}$, we classify the surface into three categories (low/medium/high albedo regions) based on $a_{\vec{r}}$ at λ = 750 nm, following Yokota et al. (2011), and assign corresponding $\mathbf{p}_{\bar{r}}$, depending on its category (Fig. 10 of Yokota et al. 2011). The BRDF was obtained by Yokota et al (2011) for $5° < \alpha < 85°$, $\theta_o < 85°$, and small $\theta_1$. Thus, when $\alpha$, or $\theta_o$ is larger, we fix the BRDF value to that of $\alpha = 85°$ or $\theta_0 = 85°$, while extrapolating it in terms of $\theta_1$. This admittedly inconsistent treatment may limit the validity of the model.

For Mercury, we used recently released reflectance maps based on observations by the MESSENGER probe (Hawkins et al., 2007; Domingue et al., 2011a, 2011b) at 430, 480, 560, 630, 750, 830, 900, and 1000 nm for $a_{\bar{r}}$ (Fig. 3). We assumed the anisotropic function derived for Mercury in Domingue et al., (2011a), which is based on Hapke model (e.g. Hapke 1986).

For the Galilean moons, the map of Io based on Galileo/SSI observations (Belton et al. 1992; Geissler et al. 1999; Becker and Geissler 2005) at 404, 559, and 756 nm was available and used for $a_{\bar{r}}$ (Fig. 3). For Europa, Ganymede and Callisto, however, we were unable to obtain global reflectance maps at different bands, but only grayscale maps were available (Becker et al. 2001). The contrast in brightness that appears in the maps is the collection of observations at different photometric bands as well as post-processing. Therefore, while we also performed simulations with those data accompanied by the simple Lambert law as a first-order approximation, i.e., $\tilde{f}(\theta_0,\theta_1,\phi;\mathbf{p}_{\bar{r}}) \equiv 1$, those results are only for reference.

In the case of Mars, we used the datasets of bolometric albedo in the 300–2900 nm range obtained with MGS/TES (Christensen et al. 2001) for $a_{\bar{r}}$ (Fig. 3). Consistently with TES data processing, we assumed the Lambert law; $\tilde{f}(\theta_0,\theta_1,\phi;\mathbf{p}_{\bar{r}}) \equiv 1$.

## 3. 2. Observed Disk-integrated Photometric Data

We also checked the availability of the observed disk-integrated photometry of the bodies to be used in the study. While there are many observations of spatially unresolved images of Solar System bodies to date (for a comprehensive multi-band photometry of Solar System planets, see, e.g., Young & Irvine 1967; Irvine et al. 1968a,b), it is not always straightforward to obtain datasets comparable to the observations from an astronomical distance because of the limitation of the observable configuration. In the following, we describe the datasets useful to examine photometric variability of each body where available.

First, it is difficult to obtain the spectra of the Terrestrial Moon as if it were observed by a distant observer, because ground-based observations always involve the near side. Although far-side observations of the Terrestrial Moon include those by Galileo's SSI (Belton et al. 1992) and the Extra-Solar Planetary Observations and Characterization (EPOCh) mission (Crow et al. 2011), as well as three images of the Earth–Moon system taken from afar by NASA's Voyager and Cassini, they are instantaneous snapshots that make it difficult to derive the color variation over the rotational period.

Observations of Mercury at various configurations also tend to encounter problems because of limitations in the observable geometry. While Mallama et al. (2002) conducted the photometry of Mercury at varying orbital and rotational phases by combining both space-based and ground-based observations, the precision was not sufficient to resolve the brightness variation along the equator.

Thus, we did not consider the observed data for the Terrestrial Moon and Mercury, but rather performed simulations based on the latest global albedo maps as described above.

On the other hand, it is feasible to observe rotational variation (at near full phases) of outer bodies, including the Galilean moons and Mars, even from Earth. For the Galilean moons, we considered datasets from Millis & Thompson (1975) (MT75) among the literature of ground-based observations of Galilean moons (e.g., Stebbins & Jacobsen 1928; Harris 1961; Johnson 1971; Blanco & Catalano 1974; Morrison et al. 1974; Nash & Johnson 1979). We processed the observed UBV colors of Galilean moons listed in Tables II, III, IV, and V of MT75, following the procedure described in that paper. In particular, we obtained the photometric variation due to spin rotation at the phase angle $\alpha = 6°$ by correcting the phase angle dependence of the observed colors. For the purpose of comparison with other bodies, we interpreted the observed colors in terms of albedo (apparent albedo; see below) by comparing the observed magnitudes and the magnitude of a perfectly reflecting Lambert sphere with an identical radius.

For Mars, we obtained the dataset of multi-band photometry observed by the EPOCh mission with seven 100-nm-wide filters ranging from 300 nm to 1000 nm (see also Crow et al. 2011). The EPOCh Mars observations were conducted from Nov 20 to 21, 2009, and the phase angle was $\alpha = 37°$, corresponding to an illumination fraction of $\sim 90\%$. The subsolar and sub-observer latitudes are $\sim 5°$ and $\sim 15°$, respectively.

The observed data considered in this paper is summarized in Table 2.

## 4. RESULTS

In this section, we present the light curves obtained by the procedure described in Section 3. After discussing the apparent features of each light curve (Section 4.1), we quantitatively intercompare the averaged colors and their variability (Section 4.2), and examine the detectability of periodicity (Section 4.3).

We describe the reflectance of the bodies in terms of their apparent albedo, which is defined as the ratio of the scattered light flux from the planet divided by the flux that would be expected for a lossless Lambert sphere at the same phase (see also Qiu et al. 2003; Cowan et al. 2009). Note that the apparent albedo at the full phase is 1.5 times as large as the geometric albedo.

### 4.1. Light curves

The simulated light curves of the Terrestrial Moon and Mercury are shown in Fig. 4. Here, the horizontal axis is the spin rotational phase angle in units of radians divided by $2\pi$ (i.e., 1 corresponds to the configuration after one orbital period), starting at the point where the sub-observer point is at $0°$ ($=360°$) longitude; in Fig. 3, the sub-observer point moves from the right edge to the left edge). For the purpose of comparison with Mars, we fix the phase angle at $\alpha = 37°$. The snapshots shown at the bottom of each panel are based on the mapping products displayed in Fig. 3. The Terrestrial Moon exhibits significant variation, with one peak and one trough corresponding to the dichotomy between hemispheres. In contrast, Mercury shows only small variations, with two enhancements around 0.25 and 0.75, corresponding to the two relatively bright regions around longitudes $270°$ and 90° shown in Fig. 3.

The modeled and observed rotational variation of the Galilean moons at $\alpha = 6°$, normalized so that the average is unity for display purposes, is shown in Fig. 5. The origin of the spin rotational phase in Fig. 5 is the superior conjunction, that is, the sub-observer point coincides with the sub-Jovian point, which is 0° (= 360°) longitude in Fig. 3; the sub-Jovian point at phases 0.0–0.5 and 0.5–1.0 is in the leading and trailing hemispheres, respectively. For Io, the model and the observed data exhibit a consistent pattern. While the peak in the reflectivity around the rotational phase 0.2–0.6 can be attributed to the area of the surface covered with $SO_2$ frost at around 140°–300° in longitude, the trough around phase 0.8 corresponds to the volcanic area covered with S (seen in orange or yellow in visible light) around 72° in longitude (Fig. 3). In reality, the reflectance of $SO_2$- and S-covered areas are the most different at $\lambda \sim 400$ nm, where S strongly absorbs the light due to the band gap (Clark 1999), whereas $SO_2$ is reflective.

The primary feature of the light curves shared by Europa and Ganymede—the sinusoidal shape originating from the dichotomy between the bright leading and dark trailing hemispheres—has been reproduced by models. In the case of Ganymede, the albedo dichotomy between the new and old terrains also modulates the light curves as a secondary effect. However, minor features of modeled Ganymede light curves show inconsistency with the data. For Callisto, the trend in the simulated light curves is almost opposite to the observed photometric variation. These are probably related to the post-processing of the input mapping datasets and the lack of the detailed BRDF models, which include the peculiar phase-angle dependence of the leading and trailing hemispheres (Section 2.3.4). Because of the insufficient consistency, we will not use these two simulated light curves in the later analysis in Section 4.3.

Figure 6 displays the simulated light curves of Mars together with the observed multi-band light curves, showing a reasonable consistency; the peak corresponds to the epoch where the bright region around Tharsis ($\sim 265°$ longitude) dominates the illuminated and visible area. The variation is evident at longer wavelengths, while it is muted at short wavelengths, which is consistent with the negligible difference in reflectance between fine-grained materials and rocks (lower panel of Fig. 1).

The variation patterns of the bodies considered here are coherent among different wavelengths. Consistently, principle component analysis on these light curves resulted in one dominant eigenspectrum with a proportion of variance larger than 99% (details not shown here). As the dimensionality of eigenvectors is, in theory, the number of the distinctive surface components minus 1 (Cowan et al. 2011; Cowan & Strait 2013), these results show no indication of the characteristic surface components more than 2.

The light curves at different orbital phase angles are different because of the narrower illuminating area and the effect of the BRDF. The effects of orbital phase angle are discussed in Section 4.3 below.

**4.2. Average Spectra and Variability**

Figure 7 summarizes the rotationally averaged spectra in the phases discussed in Section 4.1 and the peak-to-trough fractional variation at each of the wavelengths (we hereinafter discuss the time variability in terms of fractional variation amplitude, which is defined as the peak-to-trough amplitude divided by the average, unless otherwise noted). The data on the Galilean moons here are based on MT75 after binning the data points over 30°. Just for reference, we also plotted the data of Earth at $\alpha = 58°$, obtained by EPOCh (Livengood et al. 2011).

While all the atmosphereless bodies show "red" colors, the wavelength-dependent variation clearly indicates the diversity of surface materials. The bodies considered here exhibit 5–50% peak-to-trough variation amplitudes, depending on the wavelengths. Those that show large amplitudes are the Terrestrial Moon, Io, and Europa, whereas Mercury, Ganymede, and Callisto all show small variations. The former group consists of bodies with relatively recent geological activities—formation of maria on the Terrestrial Moon (compared to the more ancient cratered highlands), volcanism on Io, resurfacing on Europa—and thus may be identified as geologically interesting targets. On the other hand, members of the latter group commonly have ancient surfaces, where long-time exposure to space has resulted in muted signatures through impact mixing and space weathering.

In terms of the wavelength dependences, the variability of the Terrestrial Moon and Mercury is mostly independent of wavelengths. On the other hand, Mars shows large variation only at longer wavelengths, because at longer wavelengths the relatively efficient scattering by fine-grained materials is noticeable, while at short wavelengths global existence of iron oxides absorbs light.

In contrast to these bodies, the Galilean moons, in particular Io and Europa, show strong variability at short wavelengths. The variability of Io at short wavelengths likely originates from the contrast between polymeric sulfur, which strongly absorbs light at $\lambda \sim 400$ nm, and surrounding reflective frost. For the three icy bodies, the non-uniformity of the degree of interactions with surrounding plasmas and space weathering plays a major role in the variability at short wavelengths (Section 2.3).

In general, both average spectra and their variability change according to the phase angles, because of changes in the area of the illuminated and visible region, and the anisotropic nature of surface scattering (BRDF). In our model, the fractional variation amplitudes of the Terrestrial Moon at $\alpha = 6°, 45°,$ and $90°$ are 35%, 40%, and 44%, respectively (see Section 4.3 and Fig. 8 below).

**4.3. Detectability of Spin Rotation Period**

Based on the rotational variability of the bodies, we discuss the detectability of the spin rotation period though periodogram analysis. The determination of the planetary rotation period is important for at least two reasons. First, it is one of the parameters that enable us to assess the planet formation theory by comparing theoretical predictions (e.g., Kokubo & Ida 2007; Kokubo & Genda 2010). The spin rotation period may also undergo later evolution by, for example, the presence of the satellites. Second, knowing the planetary rotation period is necessary for phase folding the light curves, which could eventually allow us to map the surface and to take a closer look at localized geological features (Cowan et al. 2009, 2011; Oakley & Cash 2009; Fujii et al. 2010; Kawahara & Fujii 2010, 2011; Fujii et al. 2011; Fujii & Kawahara 2012; Fujii et al. 2013).

The observational precision necessary for spin period measurement is roughly expected to be inversely proportional to the variation amplitude; to detect the periodicity of sinusoid with amplitude $A$, observational noise would be roughly a few times smaller than $A$. To see more clearly, we estimate the signal-to-noise ratio (S/N) needed to detect the periodicity by the following procedure. The S/N here is defined by the ratio of planetary flux at each exposure (data point) to the standard deviation of the observational noise associated with each data point.

Given each theoretical (i.e., noiseless) light curve, mock datasets are created by adding artificial Gaussian noises with varying S/N. Subsequently, we perform Lomb-Scargle periodogram analysis (Lomb 1976; Scargle 1982; Press et al. 2007) on each mock dataset and discern whether the peak of the power spectrum is present within $P_{spin} \pm 0.1 P_{spin}$. If the peak is present, the false alarm probability (FAP) of the peak is computed in a randomization scheme (e.g., Murdoch et al. 1993). Namely, we create 1000 realizations by randomly rearranging the data points, while retaining the time spacing of observation, and perform the same periodogram analysis on them; the fraction of the realizations whose highest power exceeds the peak of the original mock data is regarded as FAP. The S/N that achieves FAP <0.1% is claimed as successful periodicity detection. To suppress the effect of randomness of the Gaussian noise, 100 sets of mock data are created for each S/N level. The smallest S/N with which the periodicity is successfully detected for all the 100 mock datasets is recorded as the S/N necessary for periodicity measurement.

The abovementioned procedure is performed on the simulated light curves of Moon (1000 nm), Mercury (1000 nm), Io (559 nm), Europa (based on the grayscale map but compatible with 551nm observation), and Mars (300-2900 nm) at three phase angles: $\alpha = 45°, 90°,$ and $135°$ ; we exclude Ganymede and Callisto from this analysis because of the insufficient matching between the simulation and the observed data (Fig. 5).

Panel (a) of Fig. 8 depicts the S/Ns necessary for periodicity measurement of the Moon, Io, Europa, and Mars when we assume 50 observations randomly scattered over 10 cycles (i.e., 5 data points per cycle on average) in a plane of S/N and fractional peak-to-trough variation amplitude. The results of different phase angles are represented by three different symbols (see caption). Overlaid black lines are the results of the same analysis on sinusoids: $1 + \frac{A}{2} \cos\left(2\pi \cdot \frac{t}{P_{spin}}\right)$ ($A$ is the peak-to-trough amplitude, $P_{spin}$ is the spin rotation period, and $t$ is the time) as a ruler, with varying amplitude $A$ and varying sampling frequency (100 observations, 50 observations, 25 observations scattered over 10 cycles).

According to our estimation, in most cases, the periodicity may be safely determined with 50 data points if observational noises per data point are 3-4 times smaller than the fractional variation amplitude. With S/N~10, the periodicity of relatively variable targets such as the Moon, Europa, and Mars is detectable after 50 observations. For Io at 559 nm, whose fraction variation amplitude is ~0.1, S/N~30 would be required.

To show this from a different angle, Panel (b) of Fig. 8 presents the number of observations needed for periodicity detection with given S/N level, assuming that sampling frequency corresponds to 5 data points per cycle on average. Observations with S/N~10 and 20 may identify periodicity after ~50 and ~30 observations, respectively, if the target is variable with fractional amplitude 0.3-0.4, like the Moon, Europa, and Mars.

Precisely speaking, spin period measurements are affected by the phase angle. At larger phase angles, local features are traced more sharply because the illuminated and visible area is narrow. As a result, the variation amplitudes of the light curves tend to become larger, and smaller-scale fluctuations start to appear on the light curves. The former helps the determination of the spin rotation period, while the latter could disturb it by accumulating the power at higher-order periodicity (e.g., $t = P_{spin}, \frac{1}{3}P_{spin}$) instead of the exact spin rotation period ($t = P_{spin}$). This is the case for Io and Mars, where larger S/N is required to locate the correct periodicity at larger phase angles. For Mercury and Mars at $\alpha = 135°$, the peak at $t = \frac{1}{2}P_{spin}$ tends to be stronger than the one at $t = P_{spin}$ because of the double-peaked light curves (Figs. 4 and 6), and results are not displayed in Fig. 8, although half period ($t = \frac{1}{2}P_{spin}$) may be detected with S/N~20. In reality, observations of exoplanets at large phase angles with high S/N will be challenging because of the much fainter planetary signal; for instance, a planet at $\alpha = 135°$ is 0.15 times as bright as that at $\alpha = 90°$ assuming a Lambert sphere. This being taken into account, it will be more feasible to search for periodicity of exoplanets at smaller phase angle.

## 5. DISCUSSION

### 5.1 Effects of Atmosphere

The aim of the present study was to lay the groundwork for future characterization of the surface of solid exoplanets through direct imaging observations. While we considered bodies smaller than Earth with minor atmospheres, the early-stage targets of future investigations are likely to be larger planets, which tend to have atmospheres. In this section, we discuss the effects of atmospheres.

In general, the presence of an atmosphere weakens the spectral signatures of the surface. In particular, a global cover of clouds or hazes (as on Venus and Titan) would drastically modify the photometric and spectroscopic signatures of the planets by efficiently scattering back the incident light well above the surface (Fujii et al. 2011). In the case of Earth, the observed multi-band photometry of Earth (with patchy cloud cover) shows ~20% fluctuation (Livengood et al. 2011), while the cloudless Earth would produce ~100% peak-to-trough variation in one spin rotation in the 700–800 nm band, due to the heterogeneous surface compositions (Ford et al. 2001).

In addition, atmospheric molecules affect the spectra through molecular absorption bands and overall Rayleigh scattering. The signatures of the surface may be seen most easily at longer wavelengths (where the effects of Rayleigh scattering are smaller) through atmospheric windows, if the effects of clouds and hazes are not significant.

Obviously, the diversity of rocky/icy exoplanet surface environments will be greater than presently expected, and future targets may or may not have atmospheres. In that sense, photometric variability associated with surface geology will be worth looking for, and such signatures would potentially work in synergy with the spectroscopic information of the atmosphere for constraining the surface environment as a whole.

### 5.2 Other Minor Bodies in the Solar System

Among the atmosphereless bodies in the Solar System we did not examined in this study, an intriguing example in terms of albedo features would be Iapetus, Saturn's third largest Moon in a synchronous orbit. One of the remarkable characteristics of Iapetus is an albedo dichotomy between the leading and trailing hemispheres: the geometric albedos of the bright and dark regions are 0.4 and 0.04, respectively (Buratti et al. 2005). One of the suggested origins of the global albedo dichotomy is that the dark materials were ejected from other satellites, possibly the outer moon Phoebe (e.g., Bell et al. 1985; Thomas & Veverka 1985; Verbiscer et al. 2009), spread into Iapetus's orbit, and precipitated onto the leading hemisphere. The sublimation of $H_2O$ ice would drastically change the distribution of the surface albedo and create the current shape of the dark region (Spencer & Denk 2010; Kimura et al. 2011). The huge albedo asymmetry of Iapetus roughly corresponds to a 160% peak-to-trough amplitude, which enables easier determination of the rotational period with S/N~5 after ~15 observations if it were an exoplanet. It is also likely that such a body "disappears" periodically, as was the case when Iapetus was first discovered by Cassini (Van Helden et al. 1984). Such a peculiar feature may be possible for synchronously rotating bodies, depending on the surrounding environment.

**5.3 Degeneracy in Estimating Surface Materials**

The variety of processes that alter surface reflection spectra suggests a high degree of degeneracy in possible surface conditions inferred from particular observed features. For instance, the albedo contrast of Earth in the "red" part of the spectrum primarily comes from the color difference of the continents and ocean, which is not the case for Mars. Then, the identification of continents/ocean should involve other diagnostics, such as the ocean's glint (e.g., Williams & Gaidos 2008, Oakley & Cash 2009, Robinson et al. 2010, 2014) or the spectroscopy of atmosphere/surface compositions (e.g., Fujii et al. 2013). It is likely a formidable task to uniquely determine the surface condition from one type of observation, and it will be important to consider multiple diagnostics available.

**6. SUMMARY**

In this study, we quantitatively examined the photometric properties of the Terrestrial Moon, Mercury, the Galilean moons, and Mars, with a focus on time variability due to spin rotation, as a basis for future exoplanet study. We simulated multi-band light curves for each body based on recent reflectance map products as well as compiled the observed data, and interpreted the observable features through the geologic characteristics of each body.

Based on the overview of the current geological insights into these bodies, we pointed out several processes that can vary the disk-averaged colors, including compositional heterogeneity due to igneous activity (on the Terrestrial Moon, Mercury, and Io), diversity in particle size (on Mars), interaction between icy surfaces and surrounding materials (Europa, Ganymede, and Callisto), and global tectonic deformation (Ganymede). These albedo traits produce rotational light curves in the UV/visible/NIR with fractional peak-to-trough variation amplitudes widely spread in the range of 5–50%. Highly variable targets may imply that the body has, or has had, major geological activity. The wavelength dependence of the variability as well as the averaged spectra are useful for clarifying the origins. The major geological processes and their effects on the disk-averaged spectra are summarized in Table 3.

If they were exoplanets, the rotational periods would in principle be observable through periodogram analysis of the light curves, depending on the S/N and the number of observations. S/N~10 per data points would allow us to detect spin rotational periods of the Terrestrial Moon, Europa, and Mars after 50 observations, while S/N~30 is needed for Io at 559 nm.

Surveying the geologic characteristics and disk-averaged photometric properties of known planets and moons illustrates realistic possibilities for investigating the surface environments of general solid exoplanets. The results presented here will also serve as a basic references to compare the known bodies with exoplanets once comparable data become available, which would be one of the approaches to determine the uniqueness or universality of Solar System bodies.

**ACKNOWLEDGEMENTS**

We gratefully acknowledge discussions with Hiroki Harakawa and Bun'ei Sato on the periodogram analysis. We also appreciate the kind support from Timothy A. Livengood and Tomohiro Usui in obtaining datasets. We are thankful to David S. Spiegel and Edwin L. Turner for helpful discussions. Insightful comments from Nicolas B. Cowan and Mark Claire significantly improved this paper. The work of Y. F. is supported from the Grant-in-Aid No. 25887024 by the Japan Society for the Promotion of Science. The authors would like to thank Enago (www.enago.jp) for the English language review.

# TABLES

**Table 1** : Data sources of global albedo maps.

| Target | Source | Effective/Central WL [nm] |
|---|---|---|
| Terrestrial Moon | SELENE/MI | 414, 749, 901, 950, 1000 |
| Mercury | MESSENGER/MDIS[1] | 430, 480, 560, 630, 750, 830, 900, 1000 |
| Io | Galileo/SSI[2] | 404 (violet), 559 (green), 756 (NIR) |
| Europa, Ganymede, Callisto | | |
| | Voyager[3] | 280-640 (clear) |
| | Galileo/SSI[3] | 611 (clear), 559 (green), 756 (NIR) |
| Mars | MGS/TES[4] | 300-2900 |

[1] http://messenger.jhuapl.edu/the mission/mosaics.html
[2] http://astrogeology.usgs.gov/search/details/Io/Voyager-Galileo/Io_Galileo_SSI_Global_Mosaic_ClrMerge_1km/cub
[3] Europa: http://astrogeology.usgs.gov/search/details/Europa/Voyager-Galileo/Europa_Voyager_GalileoSSI_global_mosaic_500m/cub
Ganymede: http://astrogeology.usgs.gov/search/details/Ganymede/Voyager-Galileo/Ganymede_Voyager_GalileoSSI_global_mosaic_1km/cub
Callisto: http://astrogeology.usgs.gov/search/details/Callisto/Voyager-Galileo/Callisto_Voyager_GalileoSSI_global_mosaic_1km/cub
[4] http://geo.pds.nasa.gov/missions/mgs/tesspecial.html

**Table 2** : Data sources for observed multi-band photometry

| Target | Source | Effective/Central WL [nm] |
|---|---|---|
| Galilean Moons | Millis & Thompson (1975) | 365 (U), 445 (B), 551 (V) |
| Mars | EPOCh[5] | 350, 450, 550, 650, 750, 850, 950 |
| Earth | EPOCh[5] | 350, 450, 550, 650, 750, 850, 950 |

[5] PDS Small Bodies Node, http://pdssbn.astro.umd.edu/index.shtml

**Table 3** : Major geological processes that vary planetary color regionally

| Process | Condition | Effect | Example |
| --- | --- | --- | --- |
| igneous activities (volcanism) | internal heat | regional cover with volcanism-related materials | Io, Terrestrial Moon |
| grained materials produced through weathering | atmosphere | brightening* | Mars |
| asymmetric interaction with surrounding plasmas/materials | tidal lock and surrounding high-density plasmas/materials | hemispheric asymmetry in color | Europa, Ganymede, (Callisto) |
| space weathering | (no/thin atmosphere) | reddening, darkening | Terrestrial Moon Mercury, Galilean moons |
| impact excavation | (no/thin atmosphere) | regional cover with ejecta deposits (many isotropic impacts could homogenize the surface) | Mercury Ganymede Callisto (Terrestrial Moon) |
| tectonic deformation | internal heat | brightening | (Ganymede) |

* --- Effects depend on the detailed properties of the soil.

**FIGURES**

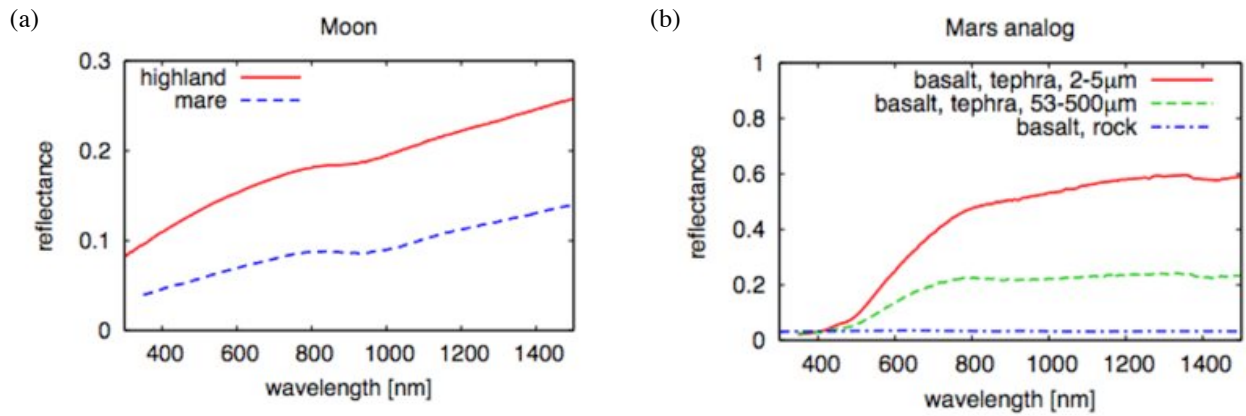

**Figure 1**   Panel (a): Reflectance spectra of two distinctive terrains, highland and mare, taken from the RELAB database (http://www.planetary.brown.edu/relab/). Samples are "62231 highland" and "12070 mare" for highland and mare, respectively. Panel (b): Reflectance Spectra of simulants of the martian surface rock, taken from the PDS Geoscience Spectral Library (http://speclib.rsl.wustl.edu/search.aspx). The datasets adopted are those entitled "Basalt; tephra/ash; altered; palagonite" with varying particle sizes (Morris et al. 2000, 2001).

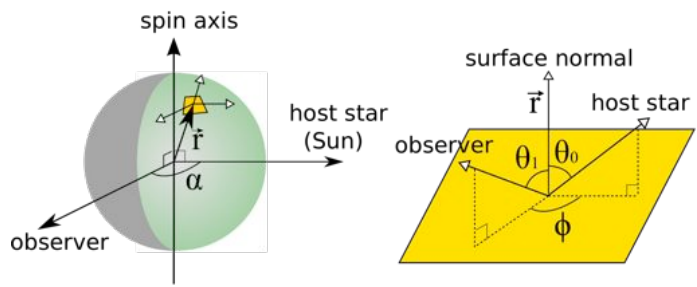

**Figure 2**   Schematic configuration of star-planet (moon)-observer system.

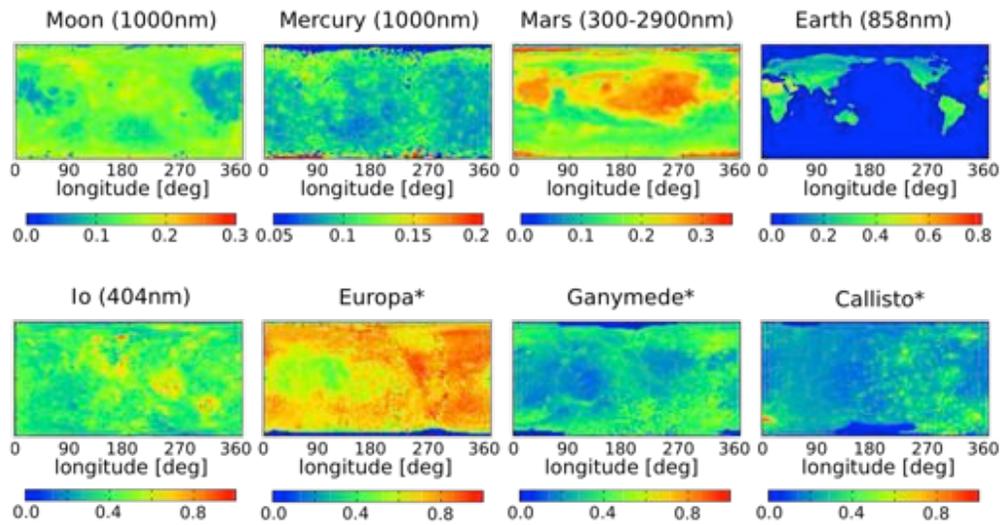

**Figure 3** Maps of planets and satellites considered in this paper, taken from the respective datasets listed in Table 1. All maps but those of Europa, Ganymede, and Callisto represent reflectance at the noted wavelengths. The data of Earth is based on MODIS BRDF/Albedo product (Schaaf et al. 2002). Maps for Europa, Ganymede, and Callisto are grayscale images based on the compilation of observations from the Voyager and Galileo spacecrafts at different filters, and include significant processing. The longitude of the terrestrial Moon, Mercury, Mars, and Earth is indicated eastward from the prime meridian. The longitude of Galilean moons is indicated eastward from the sub-Jupiter points; the left and right halves correspond to the trailing and leading hemispheres, respectively.

**Figure 4** Diurnal light curves of the terrestrial Moon and Mercury at full phase at different wavelengths. The snapshots at the bottom portray the surface albedo maps at 1000 nm in the visible area (Fig. 3).

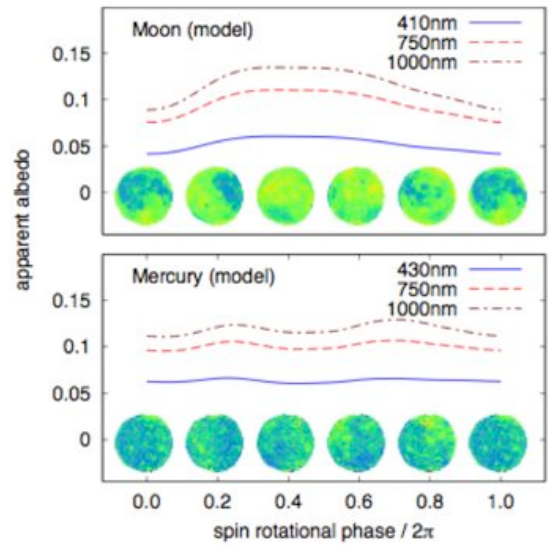

**Figure 5** Diurnal light curves of the Galilean moons. Normalized reflectance is plotted as a function of the spin rotational phase, with 0.0 being the phase where the sub-observer point is at the sub-Jovian point. The leading and trailing hemispheres are observed at 0.25 and 0.75, respectively. Points are reproduced from the UBV photometric observations of the Galilean moons by Millis & Thompson (1975). Solid and dashed lines are model light curves based on the global mapping datasets (Fig. 3); the lines for Europa, Ganymede, and Callisto are dashed to call attention to the apparent photometric inaccuracy of the input maps.

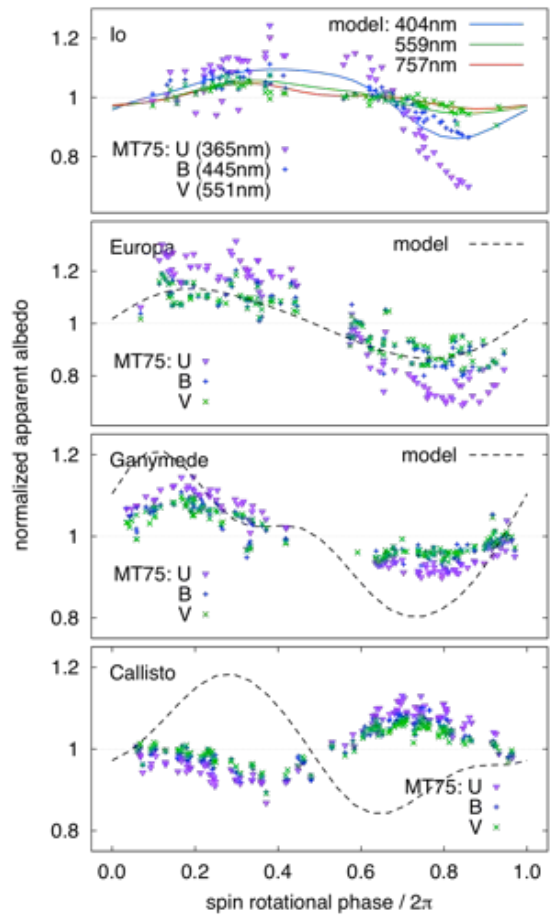

**Figure 6** Diurnal light curves of Mars at different wavelengths at phase $\alpha = 37°$. The snapshots at the bottom are based on the MGS/TES bolometric albedo map (Fig. 3). Note that the subsolar and sub-observer latitudes in the observed data are $\sim 5°$ and $\sim 15°$, respectively, while the model as well as snapshots assumes the observations from the equatorial plane for simplicity.

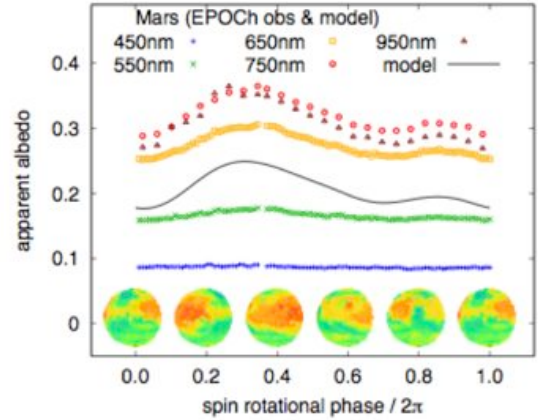

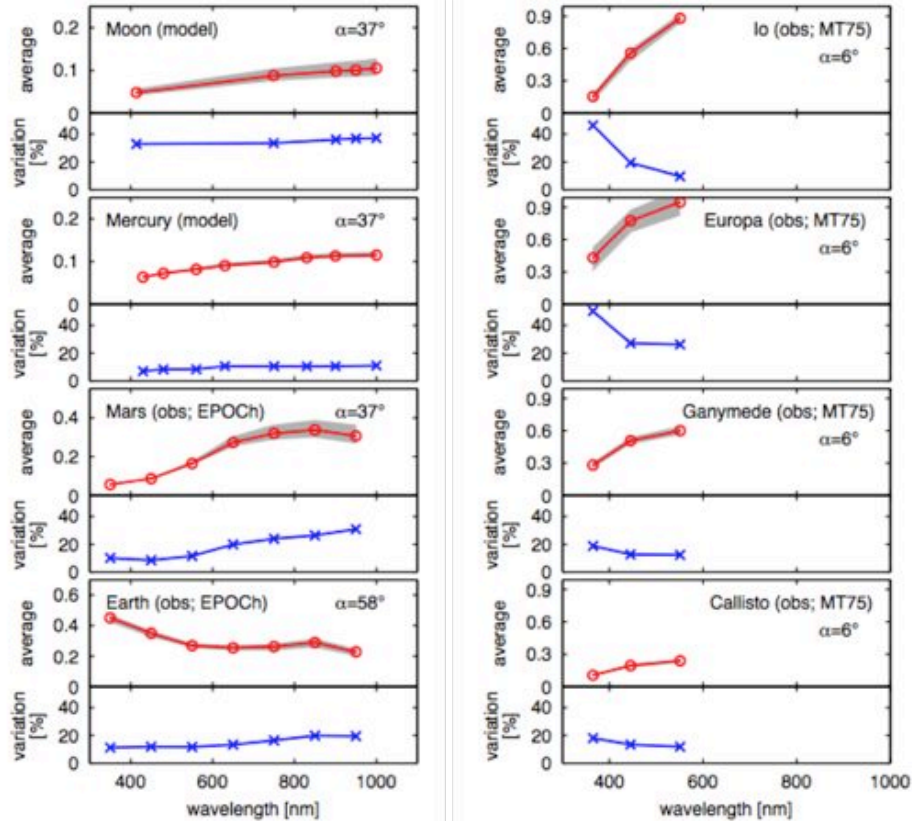

**Figure 7** Red: rotationally averaged low-resolution spectra of Solar System solid bodies. The shadowed regions show the peak-to-trough variation amplitude in one rotation. Blue: variation fraction spectra, defined as the peak-to-trough variation amplitude divided by the averaged spectra.

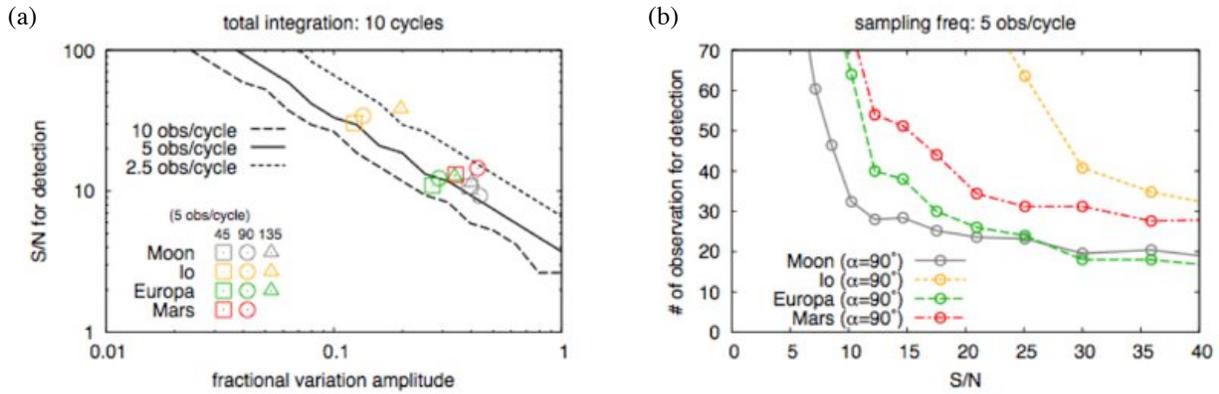

**Figure 8** Panel (a): Signal-to-noise ratio (S/N) necessary to detect the periodicity of the simulated light curves of the Terrestrial Moon at 1000 nm (gray), Io at 559 nm (orange), Europa at 551 nm (green), and Mars at 300-2900 nm (red) at three phase angle: $\alpha = 45°$ (square), $90°$ (circle), and $135°$ (triangle). Fifty observations randomly scattered over 10 cycles are assumed. For references, the S/N to detect the periodicity of sinusoids with varying amplitude and varying sampling frequency (10 obs/cycle: long-dashed, 5 obs/cycle: solid, and 2.5 obs/cycle: short-dashed) are overlaid. Panel (b): Number of observations needed to detect the periodicity with given S/N per observation. The Terrestrial Moon at 1000 nm (gray solid), Io at 559 nm (orange short-dashed), Europa at 551 nm (green long-dashed), and Mars at 300-2900 nm (red dot-dashed) located at $\alpha = 90°$ are considered. Assumed observation frequency is 5 observations per cycle on average.